# High Performance Charge Breeder for HIE-ISOLDE and TSR@ISOLDE Applications

Andrey Shornikov[1, a)], Edward N. Beebe[2], Robert C. Mertzig[1], Alexander Pikin[2] and Fredrik J. C. Wenander[1]

[1] *CERN, Geneva 23, CH-1211, Switzerland*
[2] *Brookhaven National Lab, Upton, NY 11973, USA*

a)Corresponding author: andrey.shornikov@cern.ch

**Abstract.** We report on the development of the HEC$^2$ (High Energy Compression and Current) charge breeder, a possible high performance successor to REXEBIS at ISOLDE. The new breeder would match the performance of the HIE-ISOLDE linac upgrade and make full use of the possible installation of a storage ring at ISOLDE (the TSR@ISOLDE initiative [1]). Dictated by ion beam acceptance and capacity requirements, the breeder features a 2-3.5 A electron beam. In many cases very high charge states, including bare ions up to Z=70 and Li/Na-like up to Z=92 could be requested for experiments in the storage ring, therefore, electron beam energies up to 150 keV are required. The electron-beam current density needed for producing ions with such high charge states at an injection rate into TSR of 0.5-1 Hz is between 10 and 20 kA/cm$^2$, which agrees with the current density needed to produce A/q<4.5 ions for the HIE-ISOLDE linac with a maximum repetition rate of 100 Hz. The first operation of a prototype electron gun with a pulsed electron beam of 1.5 A and 30 keV was demonstrated in a joint experiment with BNL [2]. In addition, we report on further development aiming to achieve CW operation of an electron beam having a geometrical transverse ion-acceptance matching the injection of 1$^+$ ions (11.5 μm), and an emittance/energy spread of the extracted ion beam matching the downstream mass separator and RFQ (0.08 μm normalized / ± 1% ).

## HIE-ISOLDE FACILITY OVERVIEW

After more than 20 years of successful operation of ISOLDE at the PS Booster, CERN's radioactive ion beam (RIB) facility is undergoing another major upgrade. The improvement program, called HIE-ISOLDE (where HIE stands for High Intensity and Energy), includes the study of an intensity upgrade of the radioactive ion production stage, beam quality improvement and construction of a higher energy superconducting post-accelerating linac. An overview of the ongoing program has been given in ref. [3]. Part of the beam quality improvement concerns the charge breeder, which should ideally suit the new superconducting linac and provide a beam with improved time structure to the users.

An independent parallel proposal related to post-accelerated RIBs at ISOLDE is the relocation of the Test Storage Ring (TSR) from MPIK (Heidelberg, Germany) to ISOLDE [1]. This project would enable a wide range of new experiments, thanks to a unique combination of an ISOL-facility and a heavy ion storage ring. The interfacing of TSR and the HIE-ISOLDE linac will put several stringent constraints on the RIB injection system and in particular on the charge breeder.

At ISOLDE the radioactive ions are produced continuously at the target-ion-sources. Extracted ions are magnetically mass separated and injected into a Penning-type ion trap called REXTRAP. The ion beam is accumulated in the trap and the beam is cooled using buffer-gas and RF excitation schemes [4]. The minimum cooling time is 20 ms, but in case a longer breeding time is required, or the injection rate into TSR is lower, the ions will be accumulated and kept in REXTRAP for longer time. After accumulation the ions are transferred to the charge breeder in a 10 μs bunch, and, owing to pulsed beam structure the injection efficiency is higher than for continuous ion injection. In this scheme the pulse intensity and requested acceptance of the EBIS are defined by the trapping capacity and emittance of the REXTRAP. The transverse beam emittance out of the charge breeder should fit into the acceptance of the downstream mass separator and RFQ. In particular for Very Highly Charged Ions (VHCI) it is desirable to separate the beam accumulation from the charge breeding due to long breeding times. Hence the Penning trap will be retained also in the future accelerator scheme as presented in Fig. 1.

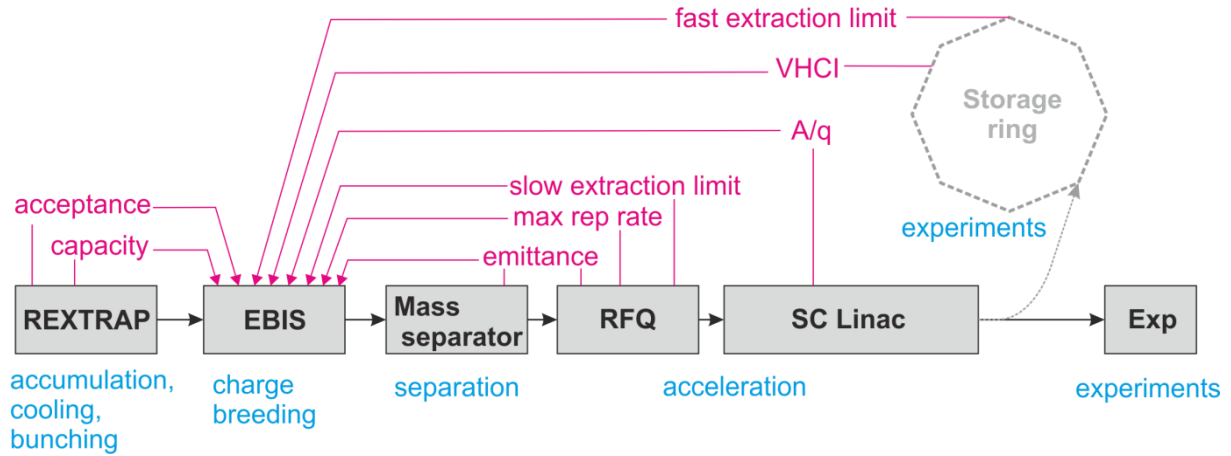

**FIGURE 1.** Scheme of the re-acceleration branch of HIE-ISOLDE and proposed storage ring installed at ISOLDE. Red lines indicate the influence of up - and downstream components on the EBIS design.

The HIE-ISOLDE linac is being extended with superconducting cavities, and the linac will retain the ability to accelerate ions in the *A/q*-range of 2.5-4.5. Owing to the individually phased cavities in the superconducting part, ions with lower *A/q*-values can be accelerated to higher energies. For instance, by reducing the *A/q* from 4.5 to 2.5 the energy is increased by a factor 1.4. The main requested quality improvement by HIE-ISOLDE is an increase of the repetition rate and extension of the extraction pulse length from the EBIS. While the superconducting part of the linac is CW, the normal conducting part is still pulsed with a maximum repetition rate of 50 Hz and RF pulse length of 800 μs (present value, to be upgraded to 2 ms with HIE-ISOLDE). An extended extraction time ($t_{extr}$) will allow to use the entire RF time-window of the linac and reduce the instantaneous counting rate (*CR*) on the detector $CR=N_{extr}/t_{extr}$ where $N_{extr}$ is the number of ions extracted from the charge breeder. *CR* above $10^6$ s$^{-1}$ [5] will cause pile-up at the MINIBALL detector. Likewise, the instantaneous rate is reduced by spreading the events over more pulses, realized by an increased repetition rate.

TSR@ISOLDE requests ions with higher charge state than REXEBIS can presently produce. The suggested program includes among other things [1]: bare ions from Z=30 through Z=70 for astrophysical p-process capture, H/Li-like Cu, Sn, Tl for study of atomic effects on nuclear half-lives and, Li/Na-like Lu, U, Th for dielectronic recombination on exotic ions. Apart from these most demanding experiments TSR favors higher charge states for experiments with gas-jet collisions as the electron stripping probability is then reduced and the storage life-time increased. The argument is also valid, although less pronounced, for in-ring decay experiments without a gas-jet but with residual gas still present. A typical storage energy of the beam inside the ring is 10 MeV/u, dictated by either experimental conditions or life-time considerations if a gas-jet is present. With a maximum rigidity of 1.57 Tm for TSR, only beams with *A/q* < 3.5 can be stored [1]. Furthermore, higher charge states will also speed up the preparation of the ion beam after it has been injected into the TSR as the electron cooling time, typically varying between 0.2 and 2 s, is given as $t_{cooling} \approx 3A/q^2$.

As multi-turn ion injection into the ring will be used the extraction pulse lengths out of the charge breeder has to be shorter than 30 μs in order to not to lose ions during the injection process. The length of the injection window is inversely proportional to the transverse emittance from the breeder, thus longer extraction times are accepted if the emittance is smaller and vice versa. The injection rate into the ring is defined by its operation mode, which can vary widely with different types of experiments. A standard injection scheme for a reaction measurement using a gas-jet target inside the ring is given in Fig. 2.

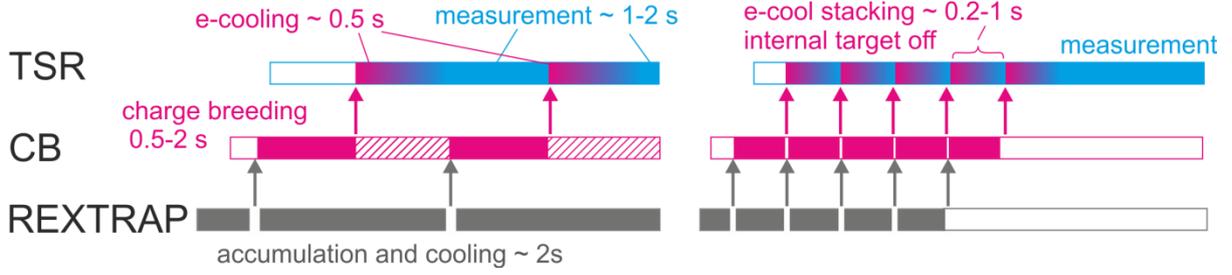

**FIGURE 2**. Examples of operation modes of TSR with RIB injection from ISOLDE.

As shown in ref. [6] the breeding of above mentioned ions requires an electron-beam current density of 10 kA/cm$^2$ and an electron energy $E_e$ up to 150 keV. Additional constraints on the charge breeder are imposed by the ion injection. In the following section the target values of the breeder design, based on the required performance and interactions with the other parts of the ISOLDE complex, are presented. Thereafter follows a report on our ongoing development and finally an outline towards an actual HEC$^2$ machine.

## CHARGE BREEDER SPECIFICATION

The full electron current is specified by the breeder capacity $N$ and the transverse ion-beam acceptance $\alpha$ requirements, given as:

$$N = f I_e L_{trap} / e v_e \tag{1}$$

$$\alpha = \frac{r_{beam}}{\sqrt{2U_{ext}}}\left[B r_{beam}\sqrt{\frac{q}{m}} + \sqrt{\frac{qB^2 r_{beam}^2}{4m} + \frac{\rho_l}{2\pi\varepsilon_0}}\right] \tag{2}$$

where $f$ is the neutralization factor, $I_e$ the electron current, $L_{trap}$ the trap length, $e$ the elementary charge, $v_e$ the electron velocity, $r_{beam}$ the beam radius, $U_{ext}$ the extraction voltage, $B$ the magnetic field, $q$ and $m$ the ion charge and mass, and $\rho_l$ the linear charge density of electrons and ions.

The capacity requirements of the new breeder should match the capacity of REXTRAP. The experimental capacity and emittance of REXTRAP measured with potassium ions using neon buffer gas are reported in [7]. As shown in Fig 3 A even though some 10$^9$ ions are injected into the REXTRAP, only about 10$^8$ can be extracted per bunch due to losses growing with the number of stored ions [7]. The number of stored ions also affects the emittance of the extracted beam (see Fig. 3 B). For high intensities the cooling becomes inefficient and the emittance of the REXTRAP is close to the emittance of the ion beam extracted from the primary target-ion-source.

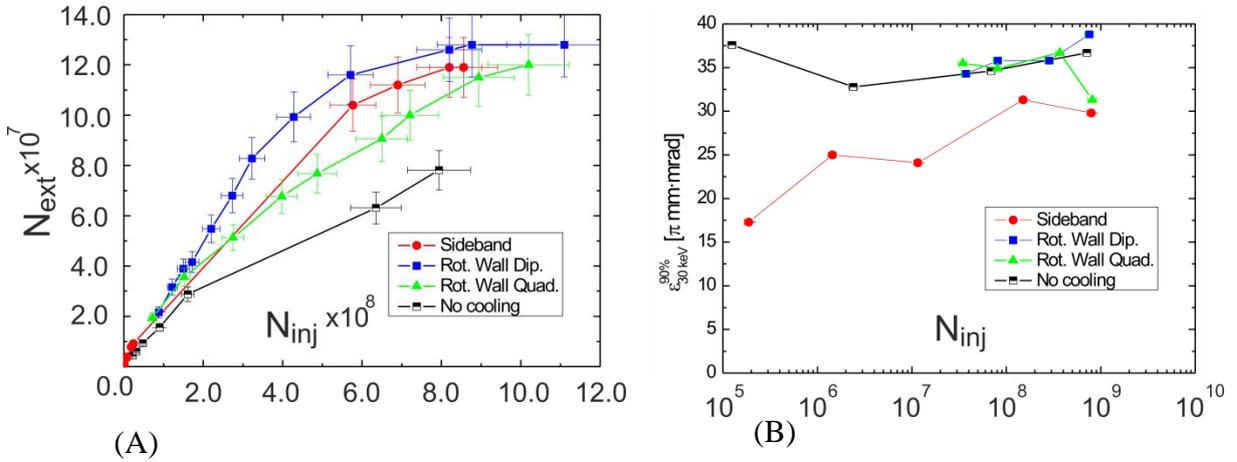

FIGURE 3. (A) Number of ions extracted from the REXTRAP as a function of number of injected ions for various ion cooling modes [7]. (B) Geometrical 90% emittance of the ion beam extracted from REXTRAP at 30 keV as a function of the number of ions injected for various ion cooling modes [7].

Such large transverse ion-acceptance is impossible to achieve with a high compression electron beam. Even REXEBIS with low compression beam at typical operation conditions (current 0.25 A, current density 125 A/cm², electron energy 4 keV, magnetic field 2 T), has only an acceptance of 11.5 microns for ions injected at 30 keV. The pulsed ion injection, however, relaxes the injection condition and allows trapping of ions even if they are not injected within the acceptance of the electron beam giving 100% overlap of the electron beam and ion cloud. A realistic goal for the ion acceptance of the upgraded high compression electron beam is a value similar to the acceptance obtained with the REXEBIS immersed low-current beam.

For the capacity estimation we consider the worst case scenario of 89+ ions, confined by a 150 keV electron beam with a neutralization factor of 0.1. In the present design version we consider $L_{trap}$ of 1 m, compared to earlier suggested 0.7 m. As shown in Fig. 4 A a current of 3.5 A will be sufficient to achieve the requested capacity even with an electron beam energy of 150 keV if we assume a trapping region of 1 m. Increasing the length further is technically challenging and the space limitations on the mechanical platform of the charge breeder prohibits a total length of breeder exceeding 4 m. A 3.5 A beam also fulfils the acceptance requirements for the requested current density of about 10 kA/cm² (see Fig. 4 B).

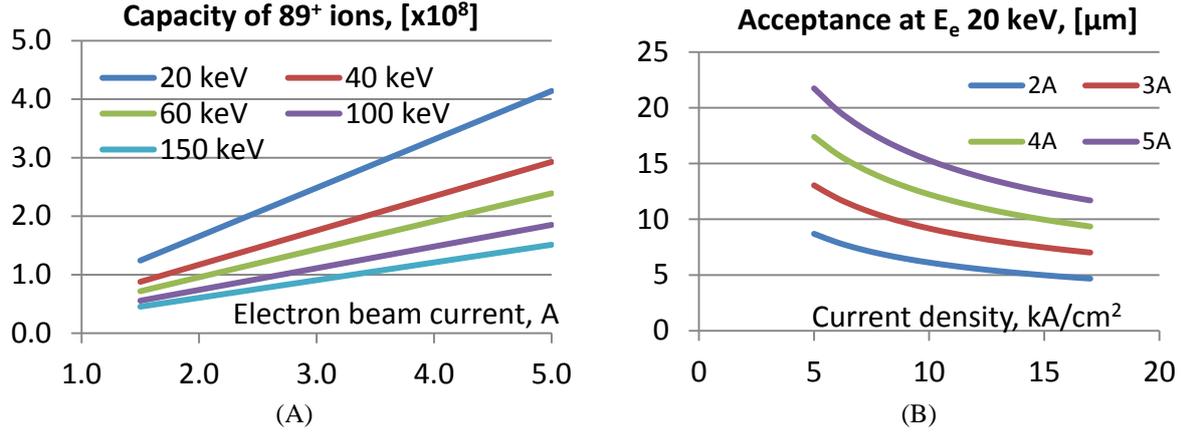

**FIGURE 4**. (A) Capacity of an EBIS for 89+ ions (x10⁸) as a function of electron beam current for various electron beam energies. (B) Transverse ion beam acceptance of an EBIS (μm) as a function of the electron beam current density for various electron beam currents at 20 keV energy.

The neutralization factor in our calculations is chosen quite conservatively. To estimate if we also need to account for the beam neutralization by the ionized residual gas we will consider two cases. The first case is fast breeding of ions with an electron energy $E_e$ = 20 keV and breeding time $\tau$ =10 ms. The second case is breeding of VHCI with $E_e$ = 150 keV and $\tau$ = 1 s. In both cases we assume an electron current $I_e$ = 3.5 A, $L_{trap}$ = 1 m and a base pressure $p$ = 1x10⁻¹⁰ mbar giving a residual gas density $n$ = 2.4x10⁸ particles per cubic centimeter. We assume that the residual atmosphere is composed entirely of hydrogen. As our working electron energies are much higher than ionization energy of hydrogen $E_i$ = 13.6 eV, we will use Bethe asymptotic approximation of the total ionization cross-section $\sigma_{TICS}$ :

$$\sigma_{TICS} = 4\pi a_0^2 \left[ A\frac{\ln(u)}{u} + \frac{B}{u} + \frac{C}{u^2} \right] \quad (3)$$

where $a_0$ is the Bohr radius of 5.29x10⁻¹¹ m, u=$E_e/E_i$, and A, B and C are Bethe coefficients for hydrogen equal to 0.2834, 1.2566 and -2.63 respectively [8]. This yields $\sigma_{TICS}$ of 7.94x10⁻¹⁹ and 1.24 x10⁻¹⁹ cm² for 20 and 150 keV respectively. Using these cross-sections we can calculate the production of ions as a current:

$$I_{res} = n\sigma_{TICS}I_e L_{trap} \quad (4)$$

which gives 0.67 and 0.1 nA for 20 and 150 keV respectively. Assuming all ions are trapped throughout the breeding period $\tau$, the accumulated charges will be 6.7 pC and 100 pC, corresponding to an additional neutralization of 1.6x10⁻⁴ and 6.5x10⁻³ respectively.

The calculation for the high energy case also gives an indication if the influx of ions from the residual gas will be sufficient for ion-ion cooling. As shown in ref. [6] breeding of U⁸⁹⁺ will be accompanied by an energy transfer of about $\Delta E$ =240 eV/q from the electron beam to the ions. The space-charge potential-well inside an electron beam of 3.5 A at 150 keV electron energy is about 140 V. To confine the ions about 8900 eV of energy must be removed from each $U^{89+}$ ion. Assuming axial potential equal to space charge potential every escaping proton removes $E_{prot}$=140 eV if ionized with the minimum potential energy on the beam axis. Accounting for random ionization location inside the electron beam will introduce a correction in the average removed energy

$$E_{prot} = 140 \text{ eV} \times \int_0^{r_{beam}} \left(\frac{r^2}{r_{beam}^2}\right) \frac{2rdr}{r_{beam}^2} = 70 \text{ eV} \tag{5}$$

where $r_{beam}$ is the electron beam radius. A total of 127 protons must be evaporated for each $U^{+89}$. If we compare it to the proton production rate calculated above (6.25x10$^8$ protons/s), we will see that the ion production from the residual atmosphere can provide enough cooling ions for about 5x10$^6$ VHCI.

For the acceptance calculations we assumed an electron beam energy of only 20 keV. This value is sufficient for the electron beam transport. After the ions have been injected with the larger ion beam acceptance, the electron energy will be ramped up to the value required by the desired ionization state (for example 150 keV as illustrated in Fig. 5 B). Fig. 5 also summarizes the period length and ion injection and extraction times for HIE-ISOLDE and TSR@ISOLDE operation.

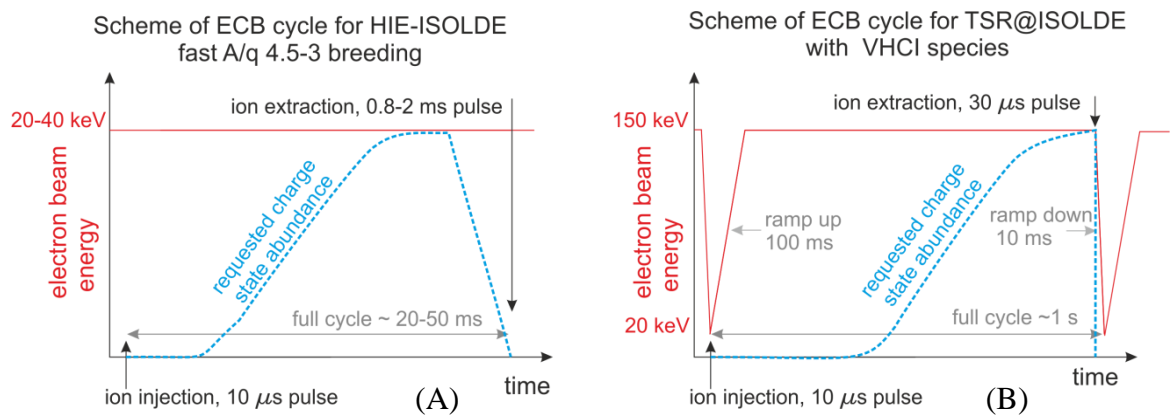

**FIGURE 5.** Charge breeding cycles of the HEC$^2$ breeder delivering beam to (A) HIE-ISOLDE (B) TSR@ISOLDE.

Breeding of VHCI has certain implications for the quality of the beam extracted out of the EBIS. Due to the above mentioned electron beam heating VHCI will receive additional energy.

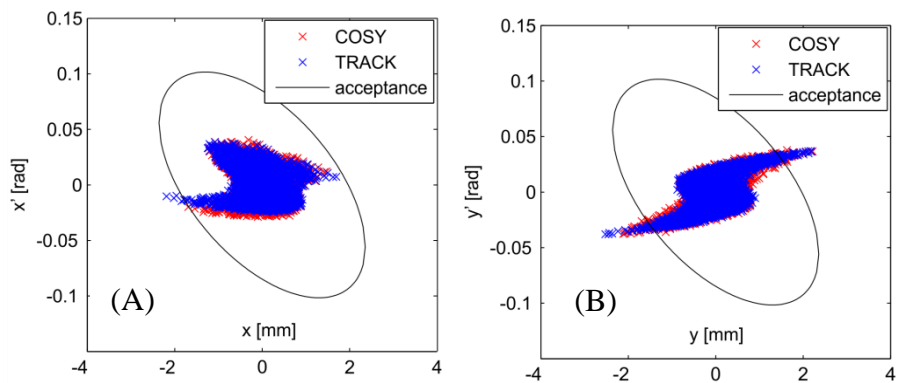

**FIGURE 6.** Phase space of an ion beam traced from the REXEBIS extraction through A/q separator and RFQ. Tracking simulations in vertical (A) and horizontal (B) planes using COSY and TRACK codes. The ion beam has a normalized emittance of 0.08 µm and an energy spread of 1% [9].

The emittance of the beam can be estimated as $\sqrt{\Delta E/U_{ext}}\, r_{beam}$. For injection into the REX-ISOLDE RFQ an energy of 5 keV/u is requested, corresponding to $U_{ext}$=13.3 kV for $U^{89+}$. For the same ion, assuming a 0.1 mm electron beam, we obtain an emittance of 13 μm (0.043 μm normalized). This is less than the 0.08 μm normalized acceptance of the downstream A/q-separator and RFQ [9]; see ion beam tracing plots and acceptance ellipses in Fig. 6.

Moreover optics simulation [9] have demonstrated that an energy spread in excess of 1% will cause beam losses. For $U^{89+}$ an energy spread of about 1.7% can be expected. Therefore, for VHCI applications ion-ion cooling will be required not only to avoid beam losses, but also to improve the extracted beam properties. Reduced ion velocities from the ion-ion cooling will also suppress charge exchange processes with neutrals inside the breeding region, especially important for VHCI.

A summary of the baseline design is given in the Table 1.

TABLE 1. Baseline design parameters of HEC$^2$ charge breeder.

| Parameter | Upgrade | Comments | REXEBIS |
|---|---|---|---|
| Electron energy [keV] | 20-150 | Ramped down for 1+ injection | 5 |
| Electron current [A] | 3.5 | | 0.25 |
| Electron current density [A/cm$^2$] | >1x10$^4$ | | 100 |
| Transverse acceptance [μm] | 11.5 | 90% non-normalized at 30 keV, pulsed injection | 11.5 |
| Ion-ion cooling needed | yes | For VHCI only | No |
| Extraction emittance [μm] | <0.07 | 1 σ, normalized at 5 keV/u | |
| Extraction time [μs] | 30 to 2000 | Variable, application specific | >50 |
| Extracted energy spread | <1% | 1 σ, at 5 keV/u | |
| Vacuum [mbar] | Low 10$^{-11}$ | Suppressing charge exchange for VHCI | Low 10$^{-11}$ |

## ONGOING DEVELOPMENT

In 2013 a high-compression prototype gun of BNL design [10] was manufactured at CERN and mounted on TestEBIS at BNL. During the first run in November 2013 the gun demonstrated a 1.5 A electron current propagating through the EBIS with an electron energy of 20-30 keV. The experimental results are given elsewhere [2]. The extracted current was limited by a 20 mA loss current on the anode. Several different reasons for the origin of this current were suggested: the current could be attributed to discharges in the gun region where crossed electric and magnetic fields (ExB) exist, electrons being reflected from the collector region, or magnetic mirror electrons unable to enter the high-field region due to excessive transverse momentum.

Discharges are likely to occur in the gap between the cathode arm and the magnetic shield, which is at the same electrical potential as the anode. The magnetic field along the cathode arm varies from near zero inside to over 1000 G at the back end of the magnetic shield (see Fig. 7 A). Thus, at some locations the field conditions for an ExB discharge are fulfilled for a wide range of anode voltages. The sparking voltage in crossed fields demonstrates a sharp drop when the magnetic field reaches a cut-off value [11] equal to:

$$B_{cutoff} = \sqrt{\frac{2m_e U}{e}}\, \frac{r_{out}}{r_{out}^2 - r_{in}^2} \tag{5}$$

where $m_e$ and $e$ are the electron mass and charge, $U$ is the applied voltage between two concentric tubes with outer and inner radii $r_{out}$ and $r_{in}$. At this cut-off value an electron emitted from the cathode will be accelerated along a cycloid trajectory which does not reach the anode due to the magnetic deflection. Such electrons can initiate an avalanche process if the vacuum conditions permit. For a 30 kV extraction voltage used in this geometry the cut-off field is 315 G (red arrow in Fig 7 A). The question whether a discharge can be sustained depends on the vacuum conditions. Experimental data on discharges in the 10$^{-7}$-10$^{-8}$ mbar vacuum region are scarce, however, based on extrapolation from lower vacuum data [12] to the range of interest, precautions against ExB discharge should be taken for the future gun if the local pressure inside the magnetic shield is in $10^{-7}$ mbar range (while the gun chamber with hot cathode is at $10^{-9}$ mbar).

Experimental studies with HEC$^2$ gun [2] also demonstrated that the ExB discharges between the anode and the vacuum chamber will significantly affect the vacuum in the gun region, even if this discharge current will not be the major contribution to the mentioned loss current. In the present configuration such discharges can be mitigated by biasing the gun chamber to the same potential as the anode.

Reflection of the electrons from the collector region was studied numerically [13]. In this study the paraxial part of the electron beam entering the collector was simulated with an increased resolution of a factor of 1400 around the beam axis. In these simulations space charge of ions produced in the collector by the incoming electron beam was neglected. This is a satisfactory assumption as protons produced in the collector and accumulated over the 2 μs drift time towards the repeller electrode biased to -2kV will constitute a charge about $8 \times 10^4$ less than that of the primary electrons in the collector, even at the base pressure of $10^{-7}$ mbar. The same calculation assuming nitrogen gives a charge 300 times less than the electron beam charge. The reflection simulation revealed a current of electrons elastically reflected back into the ionization region by the field of the extraction electrode. The reflected current cannot be reduced to zero, but optimizing the collector potentials it was reduced from 3 to 0.8 mA, out of 10 A full current. The collector settings producing the minimum amount of reflected current are similar to those used in experimental runs. Further studies demonstrated that such reflected electrons either hit the suppressor or are reflected by the magnetic mirror while re-entering the high field region. The latter return to collector and are absorbed there. Therefore this creates only a virtual loss current. Thus, it is assumed, that reflected electrons from the collector are an unlikely cause of the recorded loss current on the anode in the experiments.

Reflection of electrons with excessive transverse momentum entering the high B-field region is discussed in detail elsewhere [2]. The mechanism is under experimental study at the moment, and a temporary remedy is to increase the magnetic field in the region between the gun and main solenoid. The excessive transverse momentum is attributed to electrons being emitted from the surface of the cathode facing the Wehnelt electrode and extracted from the gap between the cathode and the Wehnelt [2]. Modifications of the gun to suppress such emission are suggested and to be implemented during the gun maintenance.

In order to measure the electron current density a time-of-flight mass spectrometer is mounted at the extraction side of the TestEBIS. First measurements are planned with ionization of the residual gas. A set of diagnostic tools including an emittance meter and an upgraded time-of-flight mass spectrometer are under commissioning at CERN, to be mounted on the TestEBIS during the scheduled maintenance.

## ROAD MAP TO THE HEC$^2$ BREEDER

Experiments at the TestEBIS provide useful opportunities to advance the HEC$^2$ project. However, due to the difference in applications the gun under study at BNL is rather a prototype for HEC$^2$ than a final solution. The main differences for the HEC$^2$ breeder are higher electron-beam energy in the trapping region and increased duty cycle (in many cases CW). At the same time the target value for the electron current is only about one third of the design current for the tested gun.

Due to an increased duty cycle issues related to electron beam losses and discharges will come into focus. Biasing of the entire gun unit to launch the electron beam at the required energy will require special attention to the discharge between the gun unit and surroundings. At the same time a lower electron current relaxes the HV conditions inside of the gun, since only a potential difference of 30 kV between cathode and anode needs to be applied.

As mentioned in the previous section, the discharges in the gun region can be mitigated by reducing the electrical potential differences. The only place where this cannot be fulfilled is the back of the gun, where also the conditions for sparking are most favorable. On the other hand the reduced cathode-anode voltage permits using a solution as given in Fig 7 B. The cathode arm can be made of isolating material such as ceramic or Macor. The heating wires and the Wehnelt bias will be guided through the middle of the cathode arm. The cathode arm is centered in the magnetic shield by an extended bushing. This bushing, extending as long as the magnetic field is present, makes the vacuum gap inside the magnetic shield equipotential. It reduces the problem of a discharge in crossed electric and magnetic fields to a problem of electric breakdown in vacuum and flashover breakdown along the isolator surface, both of which can be mitigated with other techniques.

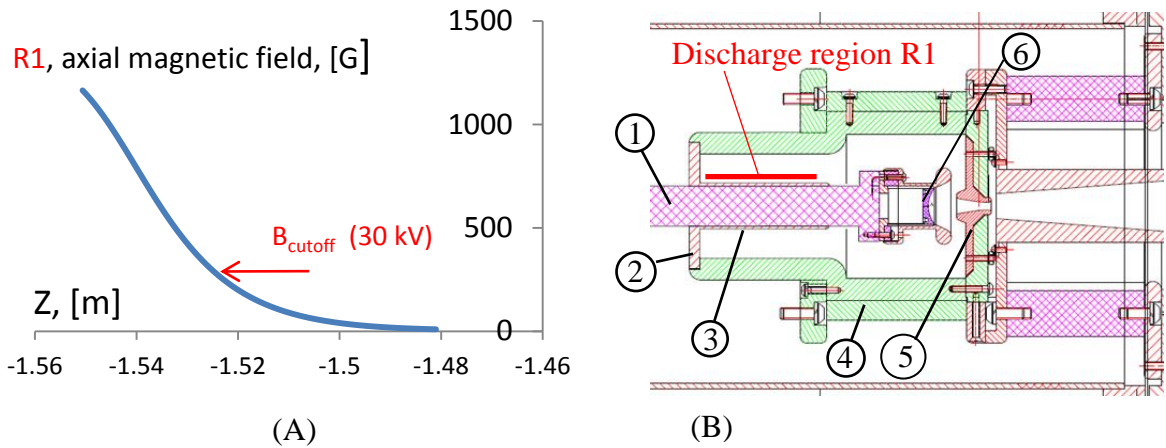

**FIGURE 7.** (A) Magnetic field along the outer surface of the cathode arm. Axial distance Z in meters from center of he trap. (B) Tentative modification of the cathode mounting system to mitigate the danger of a discharge in crossed electric and magnetic fields. The cathode (6) is mounted on an isolating ceramic support (1) fixed on the support flange (2) and covered by metal bushing (3). The support flange and the bushing are equipotential with the magnetic shield (4) and the anode (5).

In order to realize a HV biasing of the gun and its vacuum chamber with respect to the drift tube region, the following scheme is suggested (see Fig 8). Compared to the present configuration the isolation strength of the HV break will be increased from 20 to 120 kV. A gate valve will be installed between the gun chamber and the HV break to allow replacement of the gun. Due to the increased distance from the main magnet a second coil will be installed between the HV break and the main magnet to enhance the axial magnetic field strength.

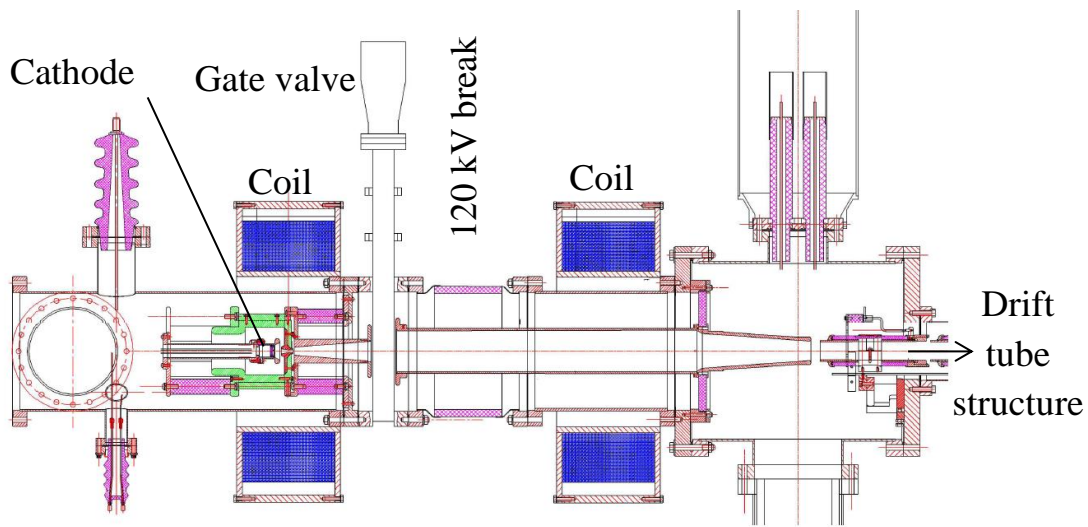

**FIGURE 8.** Tentative design of the electron gun module for the HEC$^2$ EBIS. The design assumes a biasing of the gun vacuum chamber to the anode potential, extended HV decoupling between gun and drift-tube region, and an extra magnetic coil to provide sufficient magnetic field strength for the beam transport.

The numerical simulations of the electron collector yielded several design suggestions aiming to enable CW operation with a multi-ampere beam [13].

In parallel a series of suggestions are now being implemented into the original gun design [2]. These suggestions aim to improve the electron beam quality by adjusting the beam launch optics and therefore enable operation at higher currents than hitherto achieved. Together with the HEC$^2$-specific adjustments described above it will give the first experimentally-backed conceptual design of the electron optics system for a HEC$^2$ breeder.

## ACKNOWLEDGMENTS

We would like to acknowledge financial support by the HIE-ISOLDE design study and by the CATHI Marie Curie Initial Training Network: EU-FP7-PEOPLE-2010-ITN Project number 264330. The authors want to thank Ivan Podadera-Aliseda and Matthew Fraser for their input on the REXTRAP and REX mass separator/RFQ.

## REFERENCES

1. M. Grieser et al., Eur. Phys. J. Special Topics, **207**, 1-117 (2012)
2. A. Pikin et al., "First test of BNL electron beam ion source with high current density electron beam", in *International Symposium on Electron Beam Ion Sources and Traps, EBIST'14*, AIP Conference Proceedings, this conference
3. R. Catherall et al., Nucl. Instrum. Meth. B, **317B**, 204-207 (2013)
4. F. Ames et al., Nucl. Instum. Meth. A, **538**, 17-32, (2005)
5. N. Warr et al., The Eur. Phys. J. A, **49**, 49:40 (2013)
6. A. Shornikov et al., Nucl. Instrum. Meth. B, **317B**, 395-398 (2013)
7. I. Podadera-Aliseda, "New developments on preparation of cooled and bunched radioactive ion beams at ISOL-facilities: the ISCOOL project and the rotating wall cooling", PhD thesis, CERN, Geneva, 2006
8. H. Deutsch, F. X. Bronold, and K. Becker, Int. J. of Mass Spect., **365-366**, 128-139, (2014)
9. M. A. Fraser, D. Voulot, and F. Wenander, "Beam dynamics simulations of the REX-ISOLDE A/q-separator", CERN-ACC-NOTE-2014-0017, CERN, Geneva, 2013
10. A. Pikin et al., Rev of Sci, Inst., **84**, 033303, (2013)
11. K. Wasa, *Handbook of Sputter Deposition Technology: Fundamentals and Applications for Functional Thin Films, Nano-Materials and MEMS* (William Andrew, Waltham, 2012),
12. J. M. Somerville, Proc of Phys. Soc. B, **65**, 620-629, (1952)
13. R. Mertzig et al., "Electron beam simulation from gun to collector: towards a complete solution", in *International Symposium on Electron Beam Ion Sources and Traps, EBIST'14*, AIP Conference Proceedings, this conference